\documentclass[aps,pre,onecolumn]{revtex4}
\usepackage{amsmath}
\usepackage{graphicx}

\def\eq#1{Eq.~\ref{#1}}

\def\ket#1{|{#1}\rangle}
\def\braket#1#2{\langle{#1}|{#2}\rangle}
\def\hPi#1{\hat \Pi_{#1}}

\begin{document}

\title{Microscopic mechanism for experimentally observed \\ anomalous elasticity of DNA in 2D}

\author{Nicolas Destainville, Manoel Manghi and John Palmeri}
\affiliation{Universit\'e de Toulouse; UPS; Laboratoire de
Physique Th\'eorique (IRSAMC); \\F-31062 Toulouse, France \\
CNRS; LPT (IRSAMC); F-31062 Toulouse, France}

\date{March 10, 2009}

\begin{abstract}
By exploring a recent model [Palmeri, J., M. Manghi, and N. Destainville. 2007.  \emph{Phys. Rev. Lett.} 99:088103] where DNA bending elasticity, described by the wormlike chain model, is coupled to base-pair denaturation, we demonstrate that small denaturation bubbles lead to anomalies in the flexibility of DNA at the nanometric scale, when confined in two dimensions (2D), as reported in atomic force microscopy (AFM) experiments [Wiggins, P. A., \emph{et al.} 2006. \emph{Nature Nanotech.} 1:137-141]. Our model yields very good fits to experimental data and quantitative predictions that can be tested experimentally. Although such anomalies exist when DNA fluctuates freely in three dimensions (3D), they are too weak to be detected. Interactions between bases in the helical double-stranded DNA are modified by electrostatic adsorption on a 2D substrate, which facilitates local denaturation. This work reconciles the apparent discrepancy between observed 2D and 3D DNA elastic properties and points out that conclusions about the 3D properties of DNA (and its companion proteins and enzymes) do not directly follow from 2D experiments by AFM.

\emph{Key words:} DNA; denaturation bubble; bending; AFM; wormlike chain

\end{abstract}

\maketitle

\section*{Introduction}

Whereas traditional bulk experiments provide average behaviors of dominant sub-populations, new methods exist that address DNA mechanical properties at the single-molecule level~\cite{bust,finzi,Pouget}. Observations by AFM of double-stranded DNA (dsDNA) adsorbed on a 2D substrate~\cite{hansma,Rivetti} have recently allowed a direct quantification of the distribution, $p(\theta)$, of bending angles $\theta$~\cite{vanNoort,Wiggins}. This led to the unexpected observation of an over-abundance of large $\theta$~\cite{Podgornik}, with respect to the Worm-Like Chain (WLC) model, at very short range ($\approx 5$~nm, much less than the persistence length $\approx 50$~nm).
\begin{figure}[ht]
\begin{center}
\includegraphics[width=14cm]{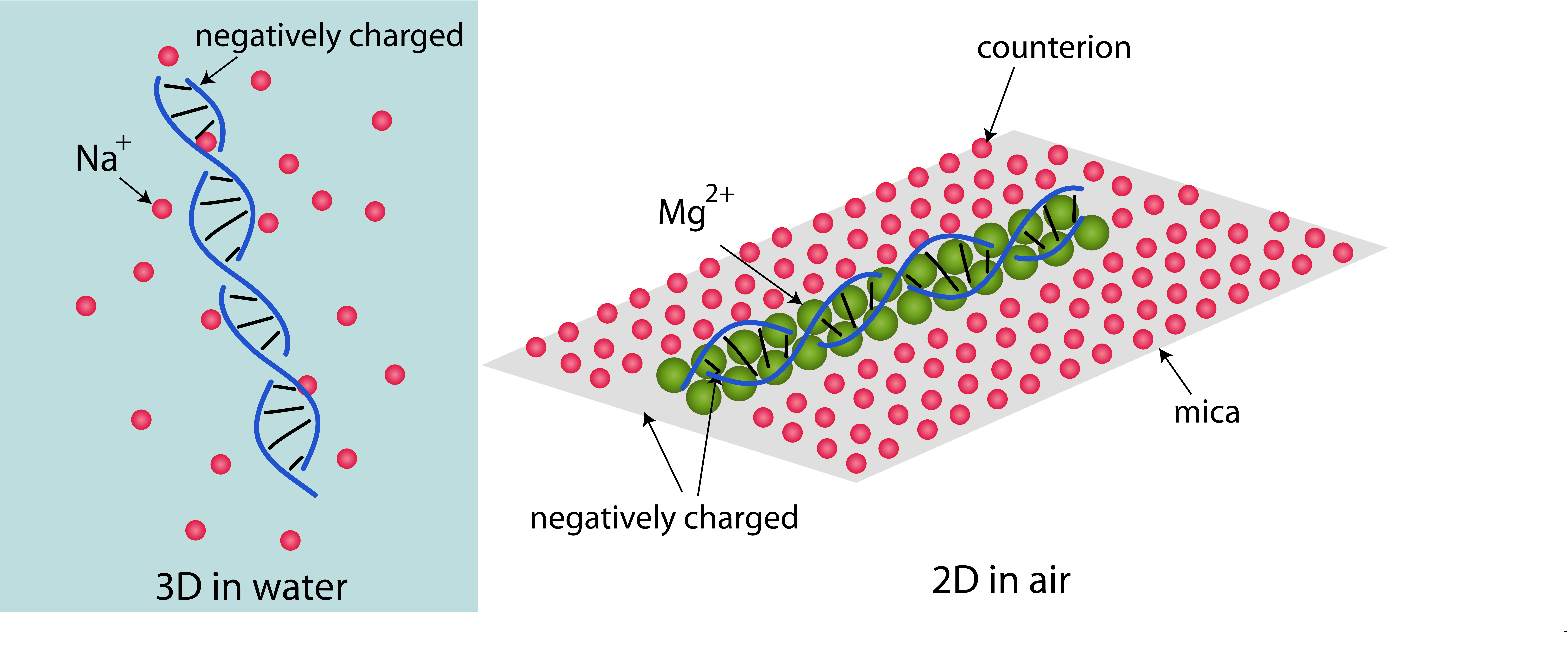}
\caption{Sketch of a dsDNA segment solvated \textit{in water} (left) with its sodium counterion cloud (the phosphate groups of the DNA backbone are negatively charged); and \textit{in air} (right), electrostatically adsorbed on a mica substrate forming an ionic crystal \textit{via} magnesium ion bridges between the DNA and the negatively charged substrate. The parameters associated with the hydrogen bonding of bps and the stacking of adjacent bases are therefore significantly modified.}
\label{sketch}
\end{center}
\end{figure}
These observations suggest that, even in the absence of any bending constraints, non-linearities, such as kinks where DNA is locally unstacked~\cite{Lankas} or small denaturation bubbles, are excited solely by thermal fluctuations with a high enough probability to be observable  at room temperature ($T_R=298.15$~K). These findings cast some doubt upon the adequacy of the WLC model traditionally adopted in 3D~\cite{kratky}. In this respect, Cloutier and Widom~\cite{Cloutier} have observed that short dsDNA, about 100 base-pairs (bp) long, formed looped complexes in 3D with a much higher probability than expected, which was attributed to partial denaturation~\cite{Yan}. However, these findings have been questioned by new experiments that pointed out a flaw in the experimental procedure~\cite{Du} and showed that short-DNA cyclization data were accurately fitted by the WLC model, without invoking kinks. 
A recent study based on flow experiments draws similar conclusions~\cite{Linna}. These converging elements are supported by all-atom numerical simulations~\cite{Lankas,Mazur} suggesting that kinks are not excited by thermal fluctuations with any measurable probability in unconstrained DNA fluctuating freely in solution.

Apart from 2D confinement, what is the difference between both types of experiments? Figure~\ref{sketch} shows a sketch of DNA fluctuating in solution or adsorbed on a mica surface as in AFM experiments~\cite{Rivetti,vanNoort,Wiggins}. These experiments are carried out in air (the solvent is dried) and DNA is electrostatically adsorbed using magnesium ions, forming an ``ionic crystal" with the charged substrate. DNA electrostatics are thus expected to be strongly affected as compared to DNA in water, hence hydrogen-bonding energies between two complementary bps and stacking energies between adjacent base aromatic rings are substantially modified.

Recently, we have proposed a solvable model where bending elasticity is intrinsically coupled to bp melting~\cite{PMD1,PMD2} in contrast to older approaches for which bending is not explicitly included~\cite{poland,wartell}. Single-stranded DNA (ssDNA) being two orders of magnitude more flexible than dsDNA, this coupling must be taken into account because local denaturation strongly increases flexibility. Here, we argue that in 2D the modification of the above denaturation parameters (bonding and stacking energies), due to adsorption, increases the probability of bp opening, which lowers in turn the bending rigidity. This explanation reconciles the apparent discrepancy between 3D and 2D experiments. 

\section*{Theory}

\subsection*{Model background}

We model dsDNA as a chain of $N$ bps $i$ ($1 \leq i \leq N$) possessing two degrees of freedom~\cite{PMD1,PMD2} : an Ising variable, $\sigma_i$, set to $+1$ (resp. $-1$) when the bp is unbroken ($U$) (resp. broken, $B$); in addition to this {\em internal} variable, an {\em external} one, the unit vector ${\bf t}_i$, sets the spatial orientation of the monomer. The Hamiltonian couples explicitly the $\sigma_i$ and ${\bf t}_i$:
\begin{equation}
H[\sigma _i,{\bf t}_i ] = \sum_{i = 1}^{N - 1}  \, \kappa(\sigma_i,\sigma_{i+1}) (1 - {\bf{t}}_{i + 1}  \cdot {\bf{t}}_i )- J\sum_{i =1}^{N - 1}  \sigma _{i + 1} \sigma _i  - \mu \sum_{i = 1}^N \sigma_i.
\label{H}
\end{equation}
The bending rigidity of the joint between bps $i$ and $i+1$, $\kappa(\sigma_i,\sigma_{i+1})$, takes different values according to the internal state of the two neighboring bps. We denote $\kappa_U \equiv \kappa(1,1)$, $\kappa_B\equiv \kappa(-1,-1)$ and $\kappa_{UB}\equiv \kappa(1,-1)=\kappa(-1,1)$. The Ising parameters $J$ and $\mu$ have the following physical meanings: $J$ is the destacking energy (energetic cost to unstack two consecutive aromatic rings); and $2\mu$ is the energy difference per bp between open and closed states.
 
This discrete WLC model coupled to an Ising one can be completely solved using a transfer matrix approach~\cite{PMD1,PMD2}. Calculating the partition function amounts to solving a {\em spinor} eigenvalue problem (formally equivalent to a quantum rigid rotator). In 3D, the orthogonal eigenstates, denoted by $| \hat\Psi_{l,m;\tau} \rangle$, are indexed by three ``quantum numbers": $l=0,1,\ldots,\infty$ and $m=-l,\ldots,l$ are the usual azimuthal and magnetic quantum numbers associated with the spatial orientation of ${\bf t}_i$ and $\tau=\pm$ is related to the ``bonding" and ``anti-bonding" bp states (as for the one-dimensional Ising model or the $H_2^+$ covalent bond). When projecting the eigenstates onto the real space basis $\ket{\sigma \Omega}$, with $\sigma$ a bp state and $(\theta,\varphi)\equiv\Omega$, the two spherical angles defining $\bf{t}$, one gets $\braket{\sigma\Omega}{\hat\Psi_{l,m;\tau}}=\psi_{l,m}(\Omega) \braket{\sigma}{l,\tau}$. The $\psi_{l,m}(\Omega) = \sqrt{4\pi} Y_{l,m}(\Omega)$, proportional to the spherical harmonics, are the eigenvectors of the pure chain model (i.e. when all $\kappa$ are set equal). The eigenvalues $\lambda_{l,\tau}$ are degenerate in $m$ and can be expressed in terms of modified Bessel functions of the first kind $I_\nu$ ($\nu={l+\frac12}$)~\cite{abra} (see Ref.~\cite{PMD2} for the expressions for the $\ket{l,\tau}$). We have $\langle {l,\tau'} | {l,\tau } \rangle  = \delta _{\tau \tau '}$, but  $\langle l',\tau ' | l,\tau  \rangle \neq \delta _{l l'}\delta _{\tau \tau '}$, because if $l \neq l'$, the matrix element is between states of different rotational symmetry. This is why our coupled model is not the trivial direct product of both the Ising and discrete WLC models.

The previous exact solution can also be found when the chain is confined to 2D, as already stated by one of us in Ref.~\cite{Palmeri}, for example when DNA is adsorbed on a substrate at thermodynamical equilibrium~\cite{Wiggins}. The spherical angles $(\theta,\varphi)$ become a single polar angle $\theta\in(-\pi,\pi]$; the spherical harmonics $\psi_{l,m}(\theta,\varphi)$ become the simpler $\psi_n(\theta)=e^{i n \theta}$, with $n$ integer; the two-dimensional analogues of the eigenvalues are denoted by $\lambda_{n,\tau}$ and the eigenvectors by $\ket{n,\tau}$~\cite{Palmeri}. 

In the model as presented here we do not take into account additional DNA degrees of freedom, such as torsion or stretching. Although we have recently demonstrated that it is possible to do so in the context of thermal denaturation~\cite{JPCM}, the additional mathematical complications of taking them into account in the calculation of the bending angle distribution would tend to obscure the basic physical mechanism leading to the onset of non-linear effective bending elasticity and is therefore not warranted here.

\subsection*{Short-distance chain statistics in 3D and 2D}

In order to compute the probability distribution $p({\bf t}_{i} \cdot {\bf t}_{i+r})$ of finding the polymer with a given relative orientation between bps $i$ and $i+r$, we introduce the partial partition function, $Z(z_i,z_{i+r})$, where all degrees of freedom are integrated out except the projections on the $z$ axis of ${\bf t}_{i}$ and ${\bf t}_{i+r}$, which are set to $ z_i$ and $z_{i+r}$ (both $\in[-1,1]$):
\begin{eqnarray}
Z (z_i,z_{i+r}) &=& \sum\limits_{\{\sigma_j=\pm 1\}}\left(\prod\limits_{j = 1}^N\int \frac{d\Omega _j}{4\pi}\right)
  \delta(\cos \theta_i - z_i) 
\delta(\cos \theta_{i+r} - z_{i+r}) \nonumber\\
&\times& \langle V|\sigma_1 \Omega_1 \rangle \prod_j \langle{\sigma_j \Omega_j |\hat P|\sigma_{j+1} \Omega_{j+1} } \rangle   \langle \sigma _N \Omega _N |V \rangle\label{Z_z},
\end{eqnarray}
where $\hat P$ is the transfer matrix and $|V \rangle$ the boundary vector~\cite{PMD1}. The complete calculation from \eq{Z_z} of $p(s)=4\pi Z (1,s)/Z$, where $s\equiv{\bf t}_{i} \cdot {\bf t}_{i+r}\equiv\cos\theta$, $\theta$ is the bending angle between two monomers separated by a distance $r$, and $Z$ is the full partition function, is given in the Supplementary Information, B. It uses the decomposition of $\hat P$ on the eigenbasis $| \hat\Psi_{l,m;\tau} \rangle$.
We have checked that boundary effects are negligible at $T_R$ as soon as $i$ is larger than a few unities. We thus give the final result for $p(s)$ in the limit of long DNA when the internal segment $[i,i+r]$ is far from both chain ends (i.e. for $N \rightarrow \infty$ and  $i \rightarrow \infty$):
\begin{equation}
p(s) = \sum_{l=0}^{\infty} \frac{2l+1}{2} P_l(s) \sum_{\tau=\pm}
\braket{0,+}{l,\tau}^2 \,e^{-r/\xi^p_{l,\tau}},
\label{bulk:limit}
\end{equation}
where $P_l(s)$ is a Legendre polynomial~\cite{abra}. \eq{bulk:limit} is  a sufficient approximation of Eq.~SI.12 for fitting purposes. This expression reveals the role of infinitely many tangent-tangent correlation lengths, $\xi^p_{l,\tau} = 1/\ln(\lambda_{0,+}/\lambda_{l,\tau})$. At $T_R$, the persistence length, $\xi^p\simeq150$~bp, coincides with the dominant correlation length $\xi^p_{1,+}$~\cite{PMD2}. \\

The same calculation holds in 2D. We find the following probability distribution (Supplementary Information, C) 
\begin{equation}
p(\theta)=\frac1{2\pi} + \frac1{\pi} \sum_{n=1}^{\infty} \cos(n \theta) \sum_{\tau=\pm}
\braket{0,+}{n,\tau}^2 \,e^{-r/\xi^p_{n,\tau}},
\label{bulk:limit2D}
\end{equation}
where $\xi^p_{n,\tau} = 1/\ln(\lambda_{0,+}/\lambda_{n,\tau})$ are also the tangent-tangent correlation lengths associated with 2D eigenmodes $\ket{n,\tau}$ with eigenvalues $\lambda_{n,\tau}$. 
For the numerical calculation of infinite series such as \eq{bulk:limit} or \eq{bulk:limit2D}, the sum is performed up to order 100 (a higher cutoff has been checked not to change numerical values).\\

At room temperature, $T_R$, one observes below (see also Fig.~\ref{dist0}a,c) that, for $\theta$ smaller than a threshold $\theta_c$, $p(s)$ and $p(\theta)$ coincide with the \textit{discrete} WLC model probability distribution, $p_{\rm DWLC}$, which is the simplified version of \eq{bulk:limit} or \eq{bulk:limit2D} when no denaturation bubbles appear (formally all $\kappa$ equal to $\kappa_U$):
\begin{eqnarray}
p_{\rm DWLC}(s) &=& \sum_{l=0}^{\infty} \frac{2l+1}{2} P_l(s) \,\left[\frac{I_{l+\frac12}(\beta\kappa)}{I_{\frac12}(\beta\kappa)}\right]^r\label{pDWLC3}\\
p_{\rm DWLC}(\theta) &=& \frac1{2\pi} \sum_{n=-\infty}^{\infty}\cos(n\theta) \left[\frac{I_n(\beta\kappa)}{I_0(\beta\kappa)}\right]^r,\label{pDWLC2}
\end{eqnarray} 
in 3D and 2D respectively (dotted lines in Figs.~\ref{dist0}a,c), with $\beta=(k_BT)^{-1}$.
In the Gaussian spin-wave approximation (GSW), $\beta\kappa\gg1$, valid here, the \textit{discrete} WLC model leads to a quadratic dependence in $\theta$. Indeed, in this case, $[I_{l+\frac12}(\beta\kappa)]^r\simeq I_{l+\frac12}(\beta\kappa/r)$. One ends up with the probability distribution for a single joint of effective bending modulus $\kappa/r$, and $p_{\rm DWLC}\simeq p_{\rm GSW}=\sqrt{\beta\kappa/(2\pi r)}\exp{[-\beta\kappa\theta^2/(2r)]}$ in 2D (see Supplementary Information, E). This implies that the free energy required to bend the polymer by an angle $\theta$ is quadratic, $F(\theta,r)=\kappa\theta^2/(2r)$. In this approximation, the bending rigidity $\kappa$ and the persistence length $\xi^p$ are related through $\xi^p=2\beta \kappa$ in 2D and $\xi^p=\beta \kappa$ in 3D~\cite{desCloiseaux}.
\section*{Results}

\begin{figure}[ht]
\begin{center}
\includegraphics*[width=\textwidth]{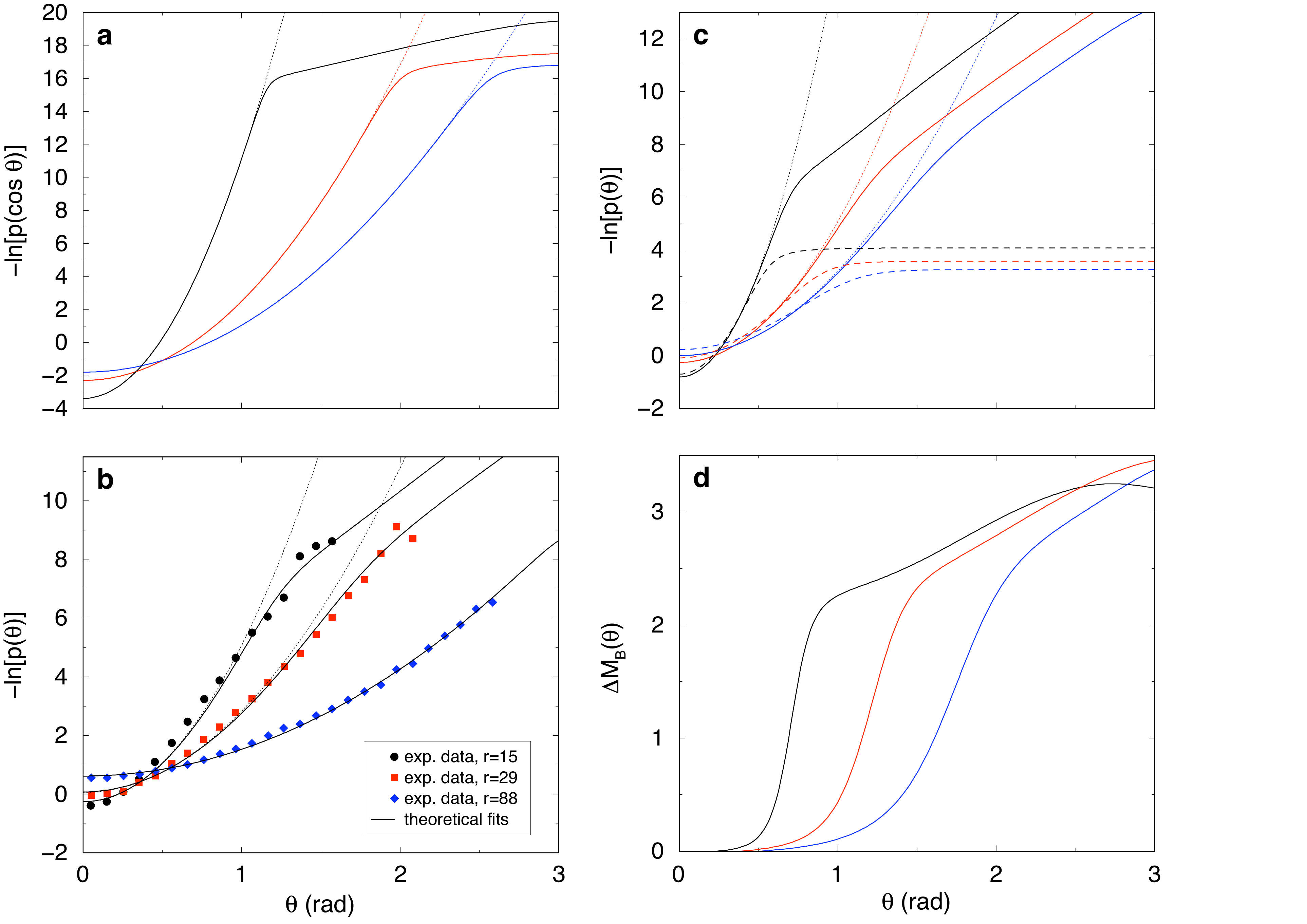}
\caption{Theoretical predictions of DNA elastic properties in two and three dimensions. {\bf (a)} Logarithm of the probability distribution $p(\cos\theta)=p(s)$ in 3D (\eq{bulk:limit}, full lines) for different values of $r=5,15,25$~bp (from left to right) compared to the WLC model (dotted lines). One bp length is $a=0.34$~nm. The Ising  and elastic parameter values (in units of $k_BT_R$) come from fits to earlier experiments~\cite{PMD1}: $\kappa_U=\kappa_{UB} = 147$; $\kappa_B = 5.54$; $\mu=1.7977$; $J=3.6674$. The probability distribution $\tilde p(\theta)$ is given by $\tilde p(\theta)= \sin\theta\, p(\cos\theta)$, because $ds=\sin\theta\, d\theta$. {\bf (b)}~Logarithm of the probability distributions $p(\theta)$ in 2D. Symbols represent experimental data taken from~\cite{Wiggins}, whereas the curves are now our best fits, from \eq{bulk:limit2D}. The curvilinear distances between monomers in~\cite{Wiggins}, namely 5, 10 and 30~nm, correspond respectively to $r=15$, 29 and 88~bp. The value $\kappa_B=5.54$ (in units of $k_BT_R$) comes from Ref.~\cite{PMD1} and $\kappa_U=160.82$ comes from fitting the $r=88$~bp set of data by a pure WLC model, as in~\cite{Wiggins} (because for such large $r$, the Gaussian character is restored). The remaining parameters $(\kappa_{UB},J,\mu)$ are fitted. One possible parameter set is $(\kappa_{UB},J,\mu)=$(20.97,1.3173,1.6685) (see Supplementary Information, D). Dotted line shown the predictions of the WLC model, for comparison.
{\bf (c)}~Logarithm of the probability distribution $p(\theta)$ in 2D. Parameter values are coming from fits (see Panel b), and $r=5,15,25$~bp (from top to bottom, full lines). Dotted line shown the predictions of the WLC model and dashed lines show the same profiles when $\kappa_B=0$. 
{\bf (d)}~Average excess chain melting $\Delta M_B(\theta)$ in 2D. Same parameter values as in Panel b. From left to right, $r=5,15,25$~bp. The elasticity is linear until a threshold $\theta_c\propto \sqrt{r}$, where excessive bending induces bp melting.}
\label{dist0}
\end{center}
\end{figure}

We first examine the distribution $p(s)\equiv p({\bf t}_{i} \cdot {\bf t}_{i+r})$ in 3D. Whereas it is dominated at large $r$ by the largest persistence length $\xi^p \simeq 150$~bp and is well described by the WLC model, this is not true at short $r$ and large $\theta$. 

Figure~\ref{dist0}a displays the probability density $p(s)$, $s={\bf t}_{i} \cdot {\bf t}_{i+r} \equiv \cos \theta$, for realistic parameters~\cite{PMD1,PMD2}. 
At $T_R$,  for $\theta$ smaller than a thrshold $\theta_c$, $p(s)$ coincides with the \textit{discrete} WLC model distribution, $p_{\rm WLC}(s)$ (\eq{pDWLC3}), the simplified version of \eq{bulk:limit} when no denaturation bubbles appear. For $\theta>\theta_c$, the plot becomes non-quadratic because of partial DNA denaturation. The threshold $\theta_c$ is estimated by equating the energetic cost of bending the polymer by an angle $\theta$ in its unmelted state, $F(\theta,r)=\kappa_U \theta_c^2/(2r)$, with the free-energy cost of nucleating a single denaturation bubble (of one bp), denoted by $\Delta G_B$, which is $\Delta G_B\simeq 17$~$k_BT$ in 3D~\cite{PMD2}. Using this scaling argument, we find
\begin{equation}
\theta_c \simeq \sqrt{\frac{2\,\Delta G_B}{\kappa_U} r}
\label{threshold}
\end{equation}
which gives a good estimate of the observed thresholds (Figure~\ref{dist0}a). The anomalies (or non-linearities) appear for larger and larger values of $\theta$ when $r$ grows, and are inexistent in the plots of $p(s)$ as soon as $r>50$~bp, i.e. at length-scales larger than $15$~nm, thus recovering standard Gaussian behavior. Indeed, setting $\theta_c=\pi$ in \eq{threshold} yields the upper limit, $r_\mathrm{max}\simeq50$~bp, as observed in the plots. This also explains why cyclization experiments with $r>50$~bp are correctly described by the WLC model~\cite{Du}. For $r<50$~bp, this local melting effect is extremely weak, occuring with a probability $\int_{\theta_c}^\pi p(\cos \theta) \sin \theta\, {\rm d} \theta \approx 10^{-7}$ for $r\geq 5$.

The situation is very different when DNA is confined in 2D. It has been demonstrated in experiments that DNA is in 2D thermodynamical equilibrium~\cite{Wiggins,Rivetti}. This is the reason why our statistical mechanical model applies and in the large $N$ limit, the probability distribution $p(\theta)$ is given by \eq{bulk:limit2D}. Plots are provided in Figs.~\ref{dist0}b,c for realistic parameter values. At large enough angles, one also sees deviations from the WLC behavior, appearing as soon as $p(\theta) \approx 0.01$~rad$^{-1}$, a now measurable value~\cite{Wiggins}. 

We fit 2D experimental data~\cite{Wiggins} in Fig.~\ref{dist0}b, using \eq{bulk:limit2D} with $\kappa_{UB}$, $J$ and $\mu$ as fitting parameters (Supplementary Information, D). The fits are good over the whole $\theta$ range.  
For the best-fit parameter sets, the fraction of melted bps for unconstrained DNA is then larger than $0.1\%$ at $T_R$, two orders of magnitude higher than in 3D~\cite{PMD1}. The predicted melting temperature, $T_m$, and transition width, both on the order of 600~K, are also much higher than their 3D analogues. Despite the high value for $T_m$ in 2D, the large transition width leads, with respect to 3D, to non-negligible bubble nucleation, even at $T_R$. In other words the loop initiation factor~\cite{poland}, $\sigma=e^{-4J_0/k_BT_R}\approx 10^{-2}$ where $J_0$ is the renormalized destacking parameter~\cite{PMD2}, is increased by several orders of magnitudes with respect to 3D~\cite{amirikyan}. The same argument as in 3D leads to $r_\mathrm{max}\simeq 120$~bp in 2D, after modifying $\Delta G_B \simeq 6.6$~$k_BT$ according to our fitted parameters. Furthermore, we display in Fig.~\ref{dist0}d the average excess of melted bps when ${\bf t}_{i} \cdot {\bf t}_{i+r}=\cos\theta$ is fixed, as compared to an unconstrained DNA (see Appendix). As anticipated, the deviation from the WLC behavior at $\theta_c$ coincides with the appearance of melted bps making the polymer more flexible.

\section*{Discussion}

How can the apparent discrepancy between 2D and 3D parameter values be explained? 
Not by the fact that the DNA used in 2D experiments are heteropolymers, whereas the values derived in 3D come from poly(dA)-poly(dT) homopolymers~\cite{PMD1}. Indeed, even for the most robust poly(dG)-poly(dC), $T_m\simeq360$~K in solution. A simple and straightforward explanation for the discrepancy in parameter values is related to the change in the DNA electrostatic energy when it is solvated in water (3D) or adsorbed through magnesium (Mg$^{2+}$) bridges on the mica in a dry environment. Indeed, it is known that slightly modifying electrostatic interactions (such as by varying the salt concentration) changes dramatically the denaturation profile of DNA in solution (see e.g.~\cite{korolev}). The energy required to break a bp, $2\mu$, and the energy to destack consecutive bps, $2J$, should also be sensitive to the change in the direct adsorption energy between mica and ds or ssDNA. Strong support for this mechanism comes from the experimental results of Wiggins \textit{et al.} themselves~\cite{Wiggins}. In their Fig.~S3, they present the angle distribution and end-to-end distance statistics for DNA adsorbed on a different-quality mica. Even though the data match to a good approximation those of their Fig. 3, a detailed analysis of the plots for $r=5$ and $7.5$~nm leads to the conclusion that the two data sets do not coincide, even taking into account error bars. This is an experimental indication that the substrate on which DNA molecules are adsorbed does indeed influence its microscopic parameters. 
Recent AFM experiments also testified to a DNA structural modification after adsorption on mica and drying~\cite{borovok}: poly(dG)-poly(dC) proves to shorten its contour length, supposedly by taking an A-DNA conformation, in contrast to poly(dA)-poly(dT) or plasmid DNA, both of which keep their B-DNA conformations.

As a result, inferring the parameters $\mu$ and $J$ from their 3D analogues is a challenging task. At the present time, the best strategy is certainly to fit them to experimental data. The above results are confirmed by recent accurate all-atom molecular dynamics simulations: Mazur has investigated in detail the short-distance angle distribution of 3D DNA and did not find any evidence for the strong deviations from a WLC distribution found experimentally in 2D~\cite{Mazur}.

Now we discuss in greater detail the role of bubble flexibility, $\kappa_B$, and of cooperativity, $J$, by comparing our model to earlier ones. In the kinkable WLC model~\cite{WigginsPRE}, kinks of vanishing rigidity can be activated by thermal fluctuations. This model and ours become physically equivalent in the $\kappa_B \rightarrow0$ limit: a 2 bp denaturation bubble plays the role of a ``kink", in the sense of a thermally activated local defect without rigidity. Our microscopic vision of a kink thus differs from Lankas \textit{et al.}'s local unstacking one~\cite{Lankas}, but yields the same short-range mechanical properties. When $\kappa_B =0$, the interesting behavior of $p(\theta)$ in the denatured region is destroyed: $p(\theta)$ becomes flat (Fig.~\ref{dist0}c), as in~\cite{WigginsPRE}, and is practically insensitive to $r$ once a kink is nucleated, because a chain segment including a kink has vanishing rigidity. This is the reason why Wiggins \textit{et al.} appeal to a different Linear Sub-Elastic Chain (LSEC) model, with a phenomenological bending energy $E_{\rm LSEC}=\Lambda |\theta|$, which enables them to satisfactorily fit their experimental data~\cite{Wiggins,LSEC}. In contrast to this LSEC model, our approach proposes a microscopic explanation  associated with bubble nucleation for the sub-harmonic behavior of $p(\theta)$. Due to excess bubble formation, our model predicts deviations from WLC (or Gaussian) behavior as soon as $r<r_{\rm max}$ with $r_{\rm max} \equiv \pi^2\kappa_U/(2\Delta G_B)$ (from \eq{threshold}). This expression differs from the LSEC model one, for which $r_{\rm max} \approx \beta \kappa_U$.

Setting $J=0$ with $\kappa_B$ finite also affects the profiles by softening the transition and increasing significantly the large angle probabilities, by a factor greater than 10 (data not shown), which confirms the importance of cooperativity (when in addition $\kappa_{UB}=0$, the model proposed in Ref.~\cite{Yan} in the context of cyclization is recovered). Neglecting $J$ or $\kappa_{UB}$ would require the use of unphysically large $\kappa_B$ values when fitting experimental data, while worsening the fit quality.

Our model is restricted to homopolymer DNA. However, a more accurate treatment should incorporate sequence effects by using bp  dependent model parameters~\cite{krueger}. Considering that the heteropolymer case is difficult to treat theoretically, and experiments provide only an average description of bending angle probability distribution, we limit ourselves here to describing the anomalous behaviour using an averaged approach. If more detailed experimental results become available, it would be worthwhile to extend our model to treat the heterogeneous case.

Currently, many AFM experiments explore DNA conformations and complexation between nucleic acids and proteins (see reviews~\cite{hansma,hansma2,cohen}). When AFM imaging is carried out on DNA~\cite{vanNoort,Wiggins,wy,dahlgren} or DNA/histone complexes~\cite{montel} in order to access their statistical and dynamical properties, effects of surface interactions on DNA structure are likely to modify sensibly these properties. More generally, our work suggests that studying DNA/companion proteins interactions by AFM~\cite{sorel,henn,wang,guo} does not provide any quantitative clue to 3D complexation.

In the cell, packaging involves wrapping DNA around positively charged histones~\cite{theCell}. It has been shown that this adsorption is mainly driven by electrostatics~\cite{oohara}. Our results suggest that in this case, DNA adsorption on a curved charged surface (such as the histone) is likely to modify profoundly local elastic and denaturation properties of dsDNA. Enhanced flexibility due to denaturation is then likely to facilitate wrapping. This mechanism might also be important for improving the accessibility of enzymes to the single strands in local bubbles~\cite{pant,sokolov} when DNA is wrapped.

\begin{figure}[ht]
\begin{center}
\includegraphics*[width=.5\textwidth]{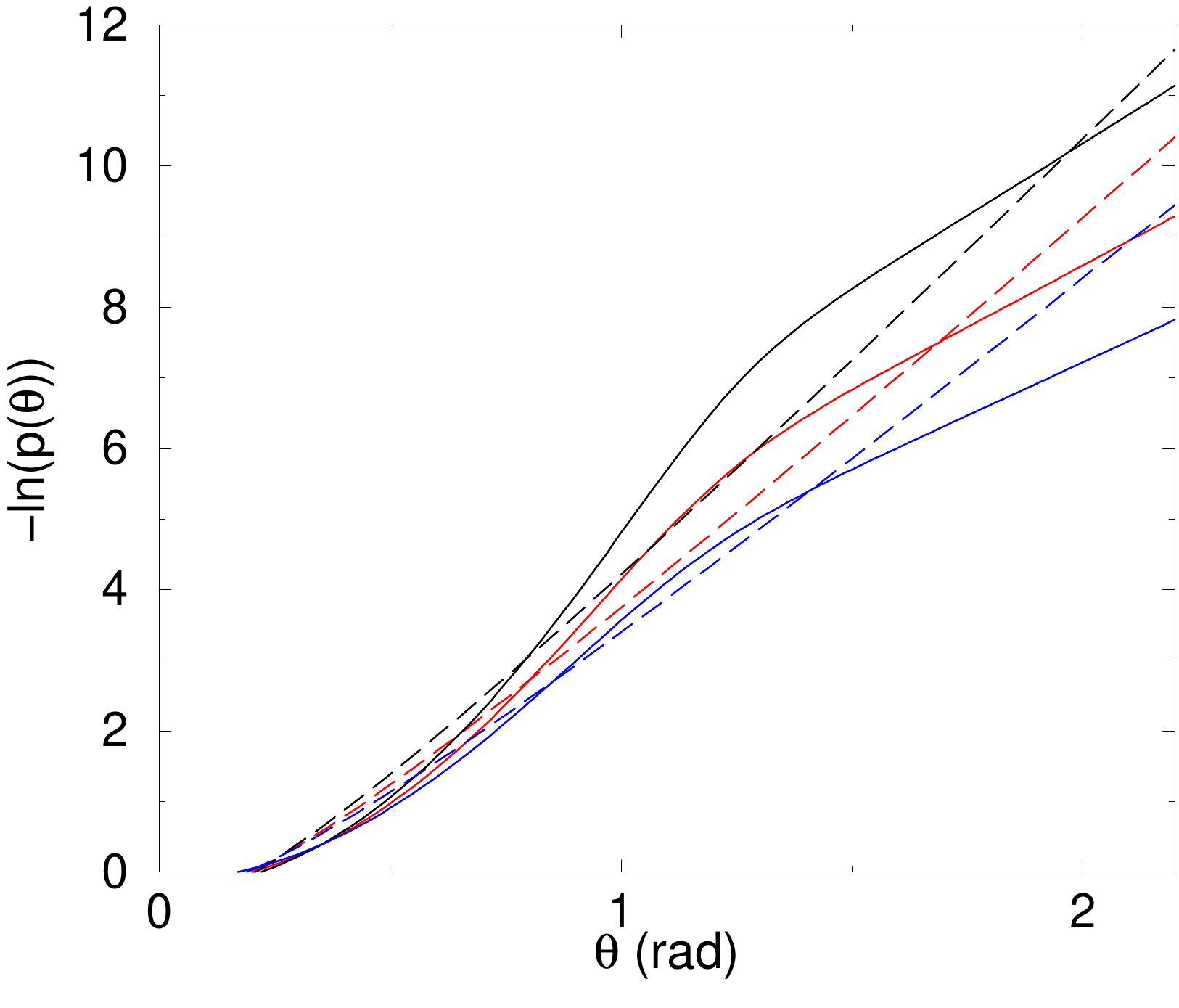}
\caption{Logarithm of the probability distributions $p(\theta)$ in 2D ($r=15$~bp), for both the LSEC model (with $\Lambda=27.2$~pN.nm~\cite{Wiggins}, dashed lines) and our theory (full lines), for increasing temperature $T$. From top to bottom, $T=298.15$~K, 330~K, and 360~K. Our model predicts that increasing $T$ enhances flexibility in a pronounced manner (note the logarithmic scale) thanks to the opening of bps. At $\theta=1.5$~rad, an experimentally accessible value~\cite{Wiggins}, the decrease of $-\ln p(\theta)$ with  increasing $T$ is twice that found with the LSEC model.
}
\label{comparison}
\end{center}
\end{figure}

One way of validating the present model at the experimental level would be to quantify the effects of temperature, which can be predicted for both our coupled model and the LSEC one~\cite{LSEC} (Fig.~\ref{comparison}; see Supplementary Information, E,  for LSEC formula). Our model predicts that increasing temperature enhances flexibility in a more pronounced manner, thanks to the opening of bps. We believe that such a deviation between the predictions of both models would be a credible experimental test of their respective validities. Additional tests of the quantitative difference between DNA properties in 3D and 2D would be to compare cyclization rates by AFM in both situations for the same dsDNA strands, or to check that denaturation remains weak in 2D when approaching the 3D melting temperature, as predicted by our results. 

\bigskip
\bigskip
\noindent {\bf Acknowledgments.}
We thank Roland R. Netz and Catherine Tardin for enlightening discussions.

\section*{Appendix: Bending-induced melting in 2D}

Following a calculation as in Ref.~\cite{WigginsPRE}, we derive the excess chain melting $\Delta M_B$ as a  function of $\theta$. It measures the average excess of melted bps in the bended chain as compared to the free, unconstrained one and is given by $\Delta M_B(\theta) \equiv -\frac{k_BT}{2} \frac{\partial}{\partial \mu} \ln p(\theta)$ (see Supplementary Information, F). The comparison of Figs.~\ref{dist0}c and d confirms that the deviation from the WLC model corresponds to the appearance of melted bps that make the polymer more flexible at short range.  An interesting feature of these calculations is the saturation of $\Delta M_B$ at a finite value, even when $r<r_{\max}$ increases. In Fig.~\ref{dist0}d, this value is close to 3, which means that the total excess number of denatured bps does not exceed 3 on average. In other words, even if $r$~bps, or more, can in principle be melted to relax the constraint ${\bf t}_{i} \cdot {\bf t}_{i+r}=\cos(\theta)$, only a few of them actually do, since it costs more energy to melt more bases, whereas, owing to the small value of $\kappa_B$, a small denaturation bubble suffices to give the whole molecule a very small resistance to torque.

\setcounter{equation}{0}
\setcounter{figure}{0}
\renewcommand{\figurename}{Fig.SI.}

\newpage

\begin{center}
{\Large \textbf{Supplementary Information to the paper
\\``Microscopic mechanism for experimentally observed anomalous
elasticity of DNA in 2D"}\\ \medskip N. Destainville, M. Manghi, J.
Palmeri}\\ \medskip Universit\'e de Toulouse; UPS; Laboratoire de
Physique Th\'eorique (IRSAMC); F-31062 Toulouse, France \\
CNRS; LPT (IRSAMC); F-31062 Toulouse, France\\ \medskip \today
\end{center}

\vspace{1cm}

\noindent\textbf{A. Effective Ising Hamiltonian}\\

Starting from the Hamiltonian $H[\sigma_i,\mathbf{t}_i]$, an effective Ising Hamiltonian, a function of the $\sigma_i$ only, is obtained by integrating out the rotational degrees of freedom ${\bf t}_i$ in the partition function, leading to a renormalized Hamiltonian,
\begin{eqnarray}
H_0[\sigma _i] &=& - \sum_{i =
1}^{N - 1} \, \left[J_0 \sigma _{i + 1} \sigma _i +\frac{K_0}2 (\sigma_{i+1} + \sigma _i )\right] - \mu \sum_{i = 1}^N
\,\sigma_i, 
\end{eqnarray}
with renormalized parameters: $J_0=J-k_BT[G_0(\beta\kappa_U)+G_0(\beta\kappa_B)-2G_0(\beta\kappa_{UB})]/4$ and $K_0=-k_BT[G_0(\beta\kappa_U)-G_0(\beta\kappa_B)]/2$ where $\beta=(k_BT)^{-1}$ and $G_0(x)=x-\ln(\sinh x/x)$ is related to the bending free energy of a single joint; $\mu$ is not renormalized (see Ref.~\cite{PMD2} for further details).

\vspace{1cm}

\noindent\textbf{B. Calculations of the probability distribution $p(\cos\theta)$ in 3D}\\

\begin{figure}[ht]
\begin{center}
\includegraphics[width=.6\textwidth]{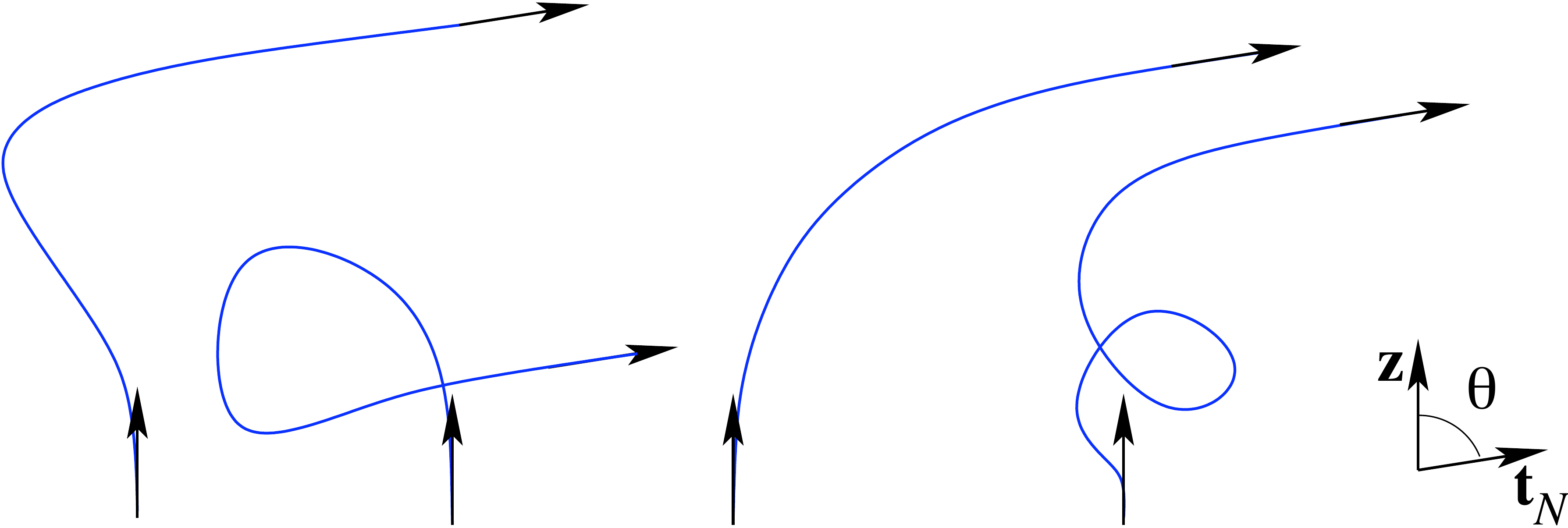}
\end{center}
\caption{Examples of chain configurations counted by the same partial partition function $Z(1,s)$ (Eq. (3) of the body of the text), in the case where the singled-out tangent vectors lie on the polymer ends, $\mathbf{t}_i=\mathbf{t}_1=\mathbf{z}$, $\mathbf{t}_{i+r}=\mathbf{t}_N$ and $s=\mathbf{z}\cdot\mathbf{t}_N=\cos\theta$.
\label{exs}}
\end{figure}

In Eq.~(3) of the body of the text, we define the partial partition function where all degrees of freedom are integrated out, except the projections on the $z$ axis of ${\bf t}_{i}$ and ${\bf t}_{i+r}$, set respectively to $z_i$ and $z_{i+r}$. Imposing the value of ${\bf t}_{i} \cdot {\bf t}_{i+r}\equiv s \in [-1,1]$ amounts to
fixing $z_i=1$ (i.e. ${\bf t}_{i}={\bf z}$), and $z_{i+r}=s$ (see Fig.SI.~\ref{exs}) and to multiplying by the solid angle $4 \pi$ to restore the rotational invariance of the whole problem with respect to ${\bf t}_{i}$, which can actually take any orientation. Thus we have $p({\bf t}_{i} \cdot {\bf t}_{i+r}=s) = 4\pi {Z(1,s)}/{Z}$ where 
\begin{equation}
Z = \langle V | \hat P^{N-1}  | V \rangle
\end{equation}
is the total partition function (see Methods and Ref.~\cite{PMD2}). 

We denote by $\hPi{z}$ the projector on the $z$ axis defined by
$\hPi{z} \ket{\sigma\Omega} =\delta(\cos \theta
-z)\ket{\sigma\Omega}$. It follows from Eq.~(3) of the body of the
text that
\begin{equation}
Z (z_i,z_{i+r})= \langle V | \hat P^{i-1} \hPi{z_i} \hat
P^{r}\hPi{z_{i+r}} \hat P^{N-r-i} | V \rangle. \label{Z_z2}
\end{equation}
In order to compute this quantity, we need to express $\hPi{z}$ in
the basis  where $\hat P$ is diagonal, namely the
$\ket{\hat\Psi_{l,m;\tau}}$. We use
\begin{equation}
\langle \sigma \Omega | \hat\Psi_{l,m;\tau} \rangle =
\psi_{l,m}(\Omega)\braket{\sigma}{l,\tau},
\end{equation}
where the spherical harmonics $\psi_{l,m}$ are defined by the
Associated Legendre polynomials $P_{l,m}$~\cite{abra} as follows:
\begin{eqnarray}
\psi_{l,m}(\Omega) & \equiv & \psi_{l,m}(\theta,\varphi)  =  k_{l,m} P_{l,|m|}(\cos \theta) e^{im\varphi}, \\
k_{l,m}&=&\sqrt{(2l+1)\frac{(l-|m|)!}{(l+|m|)!}}.
\end{eqnarray}
Below, we shall only need the Legendre polynomials $P_l(x)\equiv
P_{l,m=0}(x)$~\cite{abra}. Now we can compute
\begin{eqnarray}
 \langle \hat \Psi_{l',m';\tau'} | \hPi{z} | \hat\Psi_{l,m;\tau} \rangle
 & = & \sum_{\sigma=\pm} \int \frac{d\Omega}{4 \pi}
\langle \hat \Psi_{l',m';\tau'} | \hPi{z} | \sigma \Omega \rangle
\langle \sigma \Omega | \hat\Psi_{l,m;\tau} \rangle \\
 & = & \sum_{\sigma=\pm} \int \frac{d\Omega}{4 \pi}
\langle \hat \Psi_{l',m';\tau'} | \sigma \Omega \rangle \delta(\cos
\theta -z)
\langle \sigma \Omega | \hat\Psi_{l,m;\tau} \rangle \\
 & = & \frac{1}{4\pi} k_{l,m} k_{l',m'} \sum_{\sigma=\pm} \int_0^{2\pi} d\varphi P_{l,|m|}(z) P_{l',|m'|}(z)
e^{i(m-m')\varphi} \braket{l',\tau'}{\sigma} \braket{\sigma}{l,\tau}\\
 & = & \frac{1}{2} k_{l,m} k_{l',m} P_{l,|m|}(z) P_{l',|m|}(z)
       \braket{l',\tau'}{l,\tau} \delta_{m,m'}.
\end{eqnarray}
When $m=m'=0$, the previous equality specializes to
\begin{equation}
\langle \hat \Psi_{l',0;\tau'} | \hPi{z} | \hat\Psi_{l,0;\tau}
\rangle = \frac12 \sqrt{2l+1} \sqrt{2l'+1} P_{l}(z) P_{l'}(z)
\braket{l',\tau'}{l,\tau}.
\end{equation}
If the boundary vector $\ket{V}$ has rotational symmetry ($l=0$),
then $\braket{\hat \Psi_{l,m;\tau}}{V} = \delta_{l0}
\delta_{m0}\braket{0,\tau}{V}$. It follows that
\begin{equation}
4\pi Z (z_i,z_{i+r}) = \sum_{\{ \tau,\tau',\tau'' \}}
\sum_{l=0}^{\infty} \frac{2l+1}2
\braket{V}{0,\tau}\lambda_{0,\tau}^{i-1} P_0(z_i)
\braket{0,\tau}{l,\tau'} P_l(z_i) \lambda_{l,\tau'}^{r} P_l(z_{i+r})
\braket{l,\tau'}{0,\tau''} P_0(z_{i+r}) \lambda_{0,\tau''}^{N-r-i}
\braket{0,\tau''}{V}.
\end{equation}
This partial partition function must be compared to the complete one
\begin{equation}
Z  = \left\langle V \right|\hat P^{N - 1} \left| V \right\rangle  =
\sum_{\tau } |\langle V |0,\tau \rangle|^2 \lambda _{0,\tau}^{N - 1}
\label{Zint}
\end{equation}
in order to get
\begin{equation}
p({\bf t}_{i} \cdot {\bf t}_{i+r}=s) = \frac{ \sum_{\{
\tau,\tau',\tau'' \}}\sum_{l=0}^{\infty} (2l+1)
\braket{V}{0,\tau}\lambda_{0,\tau}^{i-1} \braket{0,\tau}{l,\tau'}
\lambda_{l,\tau'}^{r} P_l(s)\braket{l,\tau'}{0,\tau''}
\lambda_{0,\tau''}^{N-r-i}\braket{0,\tau''}{V}}{2  \sum_{\tau }
|\langle V |0,\tau \rangle|^2 \lambda _{0,\tau}^{N - 1}}
\label{p:tot}
\end{equation}
because $P_0(1)=P_0(z)=P_l(1)=1$. One can check that the
distribution is correctly normalized, $\int_{-1}^1 p(s) \; ds = 1$,
because $\int_{-1}^1 P_l(s) \; ds =2 \delta_{l,0}$.

In the limit of a long DNA where the internal segment $[i,i+r]$ is
far from both chain ends, in other words when $N \rightarrow \infty$
and then $i\rightarrow\infty$, the previous relation becomes
\begin{eqnarray}
p({\bf t}_{i} \cdot {\bf t}_{i+r}=s) & = & \sum_{l=0}^{\infty}
\frac{2l+1}{2} P_l(s) \sum_{\tau'=\pm} | \braket{0,+}{l,\tau'}|^2
\left( \frac{\lambda_{l,\tau'}}{\lambda_{0,+}}\right)^r \\
 & = & \frac12 + \sum_{l=1}^{\infty}
\frac{2l+1}{2} P_l(s) \sum_{\tau'=\pm} | \braket{0,+}{l,\tau'}|^2
e^{-r/\xi^p_{l,\tau}}.
\label{bulk:limit:SI}
\end{eqnarray}
This expression reveals the role of infinitely many correlation
lengths, the $\xi^p_{l,\tau} =
1/\ln(\lambda_{0,+}/\lambda_{l,\tau})$. At $T_R$, the persistence length $\xi^p \simeq 150$~bp coincides with $\xi^p_{1,+}$~\cite{PMD1,PMD2}. We have checked that
boundary effects are indeed negligible at room temperature
($T_R=298.15$~K) as soon as $i$ is larger than a few unities. Thus
Eq.~(\ref{bulk:limit:SI}) is a sufficient approximation of
Eq.~(\ref{p:tot}) for fitting purposes and is used in the body of the text.
Once this distribution $p(\cos\theta)$ is known, the probability distribution of $\theta$, denoted by $\tilde p(\theta)$, is simply given by $\tilde p(\theta) = p(\cos\theta)\sin\theta  {\rm d}\theta$, because $s=\cos \theta$ and $|ds|=\sin\theta \; |d\theta|$.

As a corollary, the mean value of the correlator $\langle {\bf t}_{i} \cdot {\bf
t}_{i+r} \rangle$ can be computed in this limit:
\begin{eqnarray}
\langle {\bf t}_{i} \cdot {\bf t}_{i+r} \rangle & = &  \int_{-1}^1 s \; p(s) \; ds \\
 & = & \sum_{l=0}^{\infty}
 \frac{2l+1}{2}  \left(\int_{-1}^1 s \; P_l(s) \; ds \right) \sum_{\tau'=\pm}
| \braket{0,+}{l,\tau'}|^2 \left(
\frac{\lambda_{l,\tau'}}{\lambda_{0,+}}
\right)^r \\
 & = & \sum_{\tau'=\pm}
| \braket{0,+}{1,\tau'}|^2 \left(
\frac{\lambda_{1,\tau'}}{\lambda_{0,+}} \right)^r,
\end{eqnarray}
because $\displaystyle{\int_{-1}^1 s \; P_l(s) \; ds = \frac{2}{3}
\delta_{l,1}}$. We therefore recover the result of Ref.~\cite{PMD2},
Eq.~(100), where it is pointed out that only two correlation lengths
remain in this correlator.

\vspace{1cm}

\noindent\textbf{C. Calculations of $p(\theta)$ in 2D}\\

Using the 2D algebraic background presented in the Methods and
the fact that $\langle \sigma \theta | \hat\Psi_{n;\tau} \rangle =
e^{in\theta} \braket{\sigma}{n,\tau}$, the matrix elements of the
projectors $\hPi{z}$ in the eigenbasis become
\begin{eqnarray}
\langle \hat \Psi_{n';\tau'} | \hPi{z} | \hat\Psi_{n;\tau} \rangle
  & = & \sum_{\sigma=\pm} \int_{-\pi}^{\pi} \frac{d\theta}{2 \pi}
\langle \hat \Psi_{n';\tau'} | \sigma \theta \rangle \delta(\theta
-\theta_0)
\langle \sigma \theta | \hat\Psi_{n;\tau} \rangle \\
& = & \frac{1}{2 \pi} \sum_{\sigma=\pm} e^{i(n-n')\theta_0} \braket{n',\tau'}{\sigma} \braket{\sigma}{n,\tau}\\
& = &  \frac{1}{2 \pi} e^{i(n-n')\theta_0}
\braket{n',\tau'}{n,\tau}.
\end{eqnarray}
In addition, $\braket{\hat \Psi_{n;\tau}}{V} = \delta_{n0}
\braket{0,\tau}{V}$ by the rotational symmetry of the boundary
conditions, and comparing the partial partition function, now
denoted by $Z(\theta)$, with the full one, $Z$, leads to the 2D
counterpart of Eq.~(\ref{bulk:limit:SI}):
\begin{eqnarray}
p(\theta_{r}=\theta) &=&\frac{Z(\theta)}{Z}\\
&=&\frac1{2 \pi}
\sum_{n=-\infty}^{\infty} \cos(n \theta) \sum_{\tau'=\pm} |
\braket{0,+}{n,\tau'}|^2 \left(
\frac{\lambda_{n,\tau'}}{\lambda_{0,+}}\right)^r  \label{sum}  \\
 & = & \frac1{2\pi}  + \frac1{\pi} \sum_{n=1}^{\infty} \cos(n \theta) \sum_{\tau=\pm}
\braket{0,+}{n,\tau}^2 \,e^{-r/\xi^p_{n,\tau}},
\label{bulk:limit2D:SI}
\end{eqnarray}
valid in the limit of long DNA strands, with $\theta_r \in (-\pi,\pi]$ defined by $\cos \theta_r \equiv {\bf t}_{i} \cdot {\bf t}_{i+r}$ and  $\xi^p_{n,\tau} \equiv 1/\ln(\lambda_{0,+}/\lambda_{n,\tau})$.  

\vspace{1cm}

\noindent\textbf{D. Fitting $p(\theta)$ in 2D}\\

Our first approach to fitting the 2D experimental data of Fig.~2B of the body of the text, coming from Ref.~\cite{Wiggins}, consisted in directly using the 3D parameter values, in particular those of $J$ and $\mu$, in the 2D Hamiltonian. The so-obtained angle probabilities are very far from the experimental ones, which led us to fit the model to experiment.

We display in Fig.~2B of the body of the text our best model fits, using Eq.~(\ref{bulk:limit2D:SI}) with $\kappa_{UB}$, $J$ and $\mu$, as fitting parameters. These least-square fits are good over the whole $\theta$ range.  The bp length is assumed to remain $a=0.34$~nm. Thus the curvilinear distances between monomers in~\cite{Wiggins}, namely 5, 10 and 30~nm, correspond respectively to $r=15$, 29 and 88~bp. The value (in units of $k_BT_R$) of $\kappa_B=5.54$ comes from Ref.~\cite{PMD1} and $\kappa_U=160.82$ comes from fitting the $r=88$~bp set of data by a pure WLC model, as in~\cite{Wiggins} (assuming that for such a large $r$, the Gaussian character is restored thanks to the central-limit theorem; we have checked that it is indeed the case for the parameters that we discuss now). The three remaining parameters ($\kappa_{UB}, J, \mu$) are fitted by a simulated annealing algorithm of our own. The best fit values appear to be highly degenerate, in the sense that a whole subset of parameters in the three-dimensional $(\kappa_{UB},J,\mu)$ space yields essentially the same mean-square deviation. 

This degeneracy can be related to its 3D equivalent. In Ref.~\cite{PMD2}, we have simplified the discussion by setting $\kappa_{UB}=\kappa_U$, the value for dsDNA, because varying $\kappa_{UB}$ amounts simply to changing the bare value of $J$ in order to keep the melting temperature, $T_m$, and the transition width unchanged: $J_{\rm mod} = J+\frac{k_BT}2[G_0(\beta\kappa_U)-G_0(\beta\kappa_{UB})]$~\cite{PMD2}. 
However, when it comes to the angle distribution $p$, the value of $\kappa_{UB}$ might play a more fundamental role because the different persistence lengths depend on it {\em via} the eigenvalues $\lambda$ in Eq.~(\ref{bulk:limit2D:SI}).
In 3D, we have checked that changing the value of $\kappa_{UB}$, while suitably modifying $J$, has, in practice, little influence on $p$, $N_r$, and $\Delta M_B$ (see definitions below), even at very short scales ($r=5$~bp).
In 2D, suitable values of $\kappa_{UB}$ range between 10 and 50~$k_BT_R$. Even if  $\kappa_{UB}=\kappa_U$ or $\kappa_{UB}=\kappa_B$ is held fixed, the minimization with respect to $J$ and $\mu$ alone does not give a significantly poorer fit.
Examples of parameter sets provided by simulated annealing are, in units of $k_BT_R$, $(\kappa_{UB},J,\mu)=$(20.97,1.3173,1.6685) or (45.10,0.8637,1.7885). Increasing the temperature by 20\% does not lift the degeneracy. With these values, the fraction of melted bps for an unconstrained DNA varies between $\varphi_B=0.1\%$ and 0.4\% at $T_R$. 

\vspace{1cm}

\noindent\textbf{E. Probability distribution $p(\theta)$ calculated from diverse Hamiltonians in 2D discussed in the paper}\\

We consider a pure elastic chain without bp melting, described by a 2D
single-joint Hamiltonian $H(\theta)$. Without any loss of generality, the
probability distribution $p(\theta)$ in 2D can be written as
\begin{equation}
p(\theta)=\frac{\prod_{i=1}^r\int_{-\pi}^{\pi}\,d\theta_i\,\exp\left[-\beta
H(\theta_i)\right]\,\delta(\theta-\theta_{i,i+r})}{\prod_{i=1}^r\int_{-\pi}^{\pi}\,d\theta_i\,\exp\left[-\beta
H(\theta_i)\right]}
\end{equation}
where $\theta \equiv \theta_{i,i+r}=\sum_{j=1}^{i+r-1}\theta_j$ is the bending
angle between ${\bf t}_i$ and ${\bf t}_{i+r}$. By introducing the
Fourier transform of the $\delta$ distribution, we get
\begin{equation}
p(\theta)=\frac1{2\pi} \int_{-\infty}^{\infty}\,d\omega\, e^{i
\omega\theta}\left[\frac{z(\omega)}{z(0)}\right]^r \label{fourier}
\end{equation}
where the characteristic function of the single-joint Hamiltonian is
defined by
\begin{equation}
z(\omega)=\int_{-\pi}^{\pi}\,d\theta\, \exp{[-i \omega\theta-\beta
H(\theta)]}. \label{char}
\end{equation}
We consider four cases:
\begin{enumerate}

    \item For a Gaussian Hamiltonian (Gaussian Spin Wave approximation, GSW),
    $H(\theta)=\kappa\theta^2/2$ with the approximation \mbox{$\theta\in(-\infty,\infty)$},
    we have $z(\omega)/z(0)=\exp[-\omega^2/(2\beta\kappa)]$ and we easily get the Gaussian
    probability distribution in 2D
    \begin{equation}
    p_{\rm GSW}(\theta)=\sqrt{\frac{\beta\kappa}{2\pi r}}\,\exp\left(-\frac{\beta\kappa\theta^2}{2r}\right).
    \end{equation}

    \item For the discrete wormlike chain where $H(\theta)=\kappa (1-\cos\theta)$ is periodic in $\theta\in(-\pi, \pi]$, Eq.(\ref{fourier}) reduces to a decomposition in Fourier series of modes $n$ with $z(\omega)$ becoming $e^{- \beta\kappa}I_n(\beta\kappa)$ and we find
    \begin{equation}
    p_{\rm DWLC}(\theta)=\frac1{2\pi} \sum_{n=-\infty}^{\infty}\cos(n\theta) \left[\frac{I_n(\beta\kappa)}{I_0(\beta\kappa)}\right]^r.
    \end{equation}
    This is equivalent to Eq.~(\ref{bulk:limit2D:SI}) in the simple case where all the $\kappa$ are set equal.

    \item For the LSEC model~\cite{Wiggins,LSEC}, where $H(\theta)=\Lambda|\theta|$ with the approximation $\theta\in(-\infty,\infty)$, we have $z(\omega)/z(0)=1/[1+(\omega/\beta\Lambda)^2]$ and the probability distribution is
    \begin{equation}
    p_{\rm LSEC}(\theta)=\frac{2^{\frac12-r}\beta\Lambda }{\sqrt{\pi}} \frac{(\beta\Lambda |\theta|)^{r-\frac12}K_{\frac12-r}(\beta\Lambda|\theta|)}{\Gamma(r)}\label{pLSEC},
    \end{equation}
    where $\Gamma$ is the Euler function and $K_\nu(x)$ the modified Hankel function~\cite{abra}. Equation~(\ref{pLSEC}) is plotted for different temperature values in Fig.~3 of the body of the text. Note that in the case where $\theta\in(-\pi,\pi]$, the above result has a negligible correction to $p$ on the order of $\exp(-\beta\Lambda\pi)\simeq 5\times10^{-10}$~rad$^{-1}$ if $\beta\Lambda\simeq 6.8$~\cite{Wiggins}.

 \item Both the GSW and LSEC models are special cases of a generalized model
    where $ H(\theta)=\alpha|\theta|^\eta$. With the
    approximation $\theta\in(-\infty,\infty)$, we recover the GSW model when $\eta=2$,
    $\alpha=\kappa/2$ and the LSEC model  when $\eta=1$,
    $\alpha=\Lambda$. Although the general case is harder to
    handle for arbitrary $r$, two limits, $r=1$ and large $r$, are easily studied. When
    $r=1$,
 \begin{equation}
    p(\theta)=\frac{\exp[ -\beta H(\theta)]}{\int_{-\infty}^{+\infty} d\theta' \exp[ -\beta
    H(\theta')]} = \frac{\eta (\beta \alpha)^{1/\eta}}{2\Gamma(1/\eta)} \exp[
    -\beta \alpha|\theta|^\eta], \qquad (r=1).
    \label{pshort}
 \end{equation}

    For large $r$,
    $\left[z(\omega)/z(0)\right]^r$  becomes a sharply peaked Gaussian function centered at
    $\omega=0$: $\left[z(\omega)/z(0)\right]^r \simeq \exp[-r \omega^2 \langle \theta^2 \rangle/2]$.
    In this limit $p(\theta)$ becomes effectively Gaussian,
    \begin{equation}
    p(\theta) \simeq \sqrt{\frac{\beta\kappa_{\rm eff}}{2\pi r}}\,
    \exp\left(-\frac{\beta\kappa_{\rm eff}\theta^2}{2r}\right), \qquad (\rm{large}\,\,\, r),
    \label{plong}
    \end{equation}
    where $\kappa_{\rm eff}(\alpha, \eta; \beta) \equiv 1/ (\beta \langle
    \theta^2 \rangle ) = (\beta\alpha)^{2/\eta} \Gamma(1/\eta)
    /[\beta\Gamma(3/\eta)]$. The effective bending rigidity, $\kappa_{\rm
    eff} =(\xi/\ell)\beta^{-1}$, can be written
    in terms of the asymptotic persistence length,
    $\xi$, the segment length, $\ell$, and the temperature.
    It is only for  $\eta=2$ that $p(\theta)$ is Gaussian at all
    length scales $r$. For strongly subharmonic
    models ($\eta$ well below 2), the large angle bending  distribution
    can be much higher than the Gaussian model
    prediction for $r \ll \xi$ (see Fig. 3A of
    Wiggins et al.~\cite{Wiggins} for the $\eta=1$ case). For non-Gaussian models, $p(\theta)$
    interpolates smoothly between an ``anomalous" behavior [Eq.~(\ref{pshort})]
    and an effective Gaussian one [Eq.~(\ref{plong})] as $r$ increases from one past the persistence length,
    $\xi$. For large $r$, even at large $\theta$, $p(\theta)$ is effectively Gaussian because it is determined mainly by
    those high entropy configurations for which almost all
    joint angles, $\theta_j$, are small (therefore in this case only the small $\theta$ behavior of
    $H(\theta)$ is important).

    From fitting the large $r$ data to the Gaussian model, Wiggins et al.
    found $\beta\kappa_{\rm eff} = \xi/\ell = 54\, \rm{nm}/ 2.5\, \rm{nm} \simeq 22 $, which implies that
    $ \beta\Lambda = \beta\alpha_{\rm LSEC} = (2\beta\kappa_{\rm eff})^{1/2} = 6.6$,
    close to the value of 6.8 found by fitting their LSEC model
    to experiment using Monte Carlo simulations.

\end{enumerate}

\vspace{1cm}

\noindent\textbf{F. Bending-induced melting in 2D}\\

\begin{figure}[ht]
\begin{center}
\includegraphics[width=.4\textwidth]{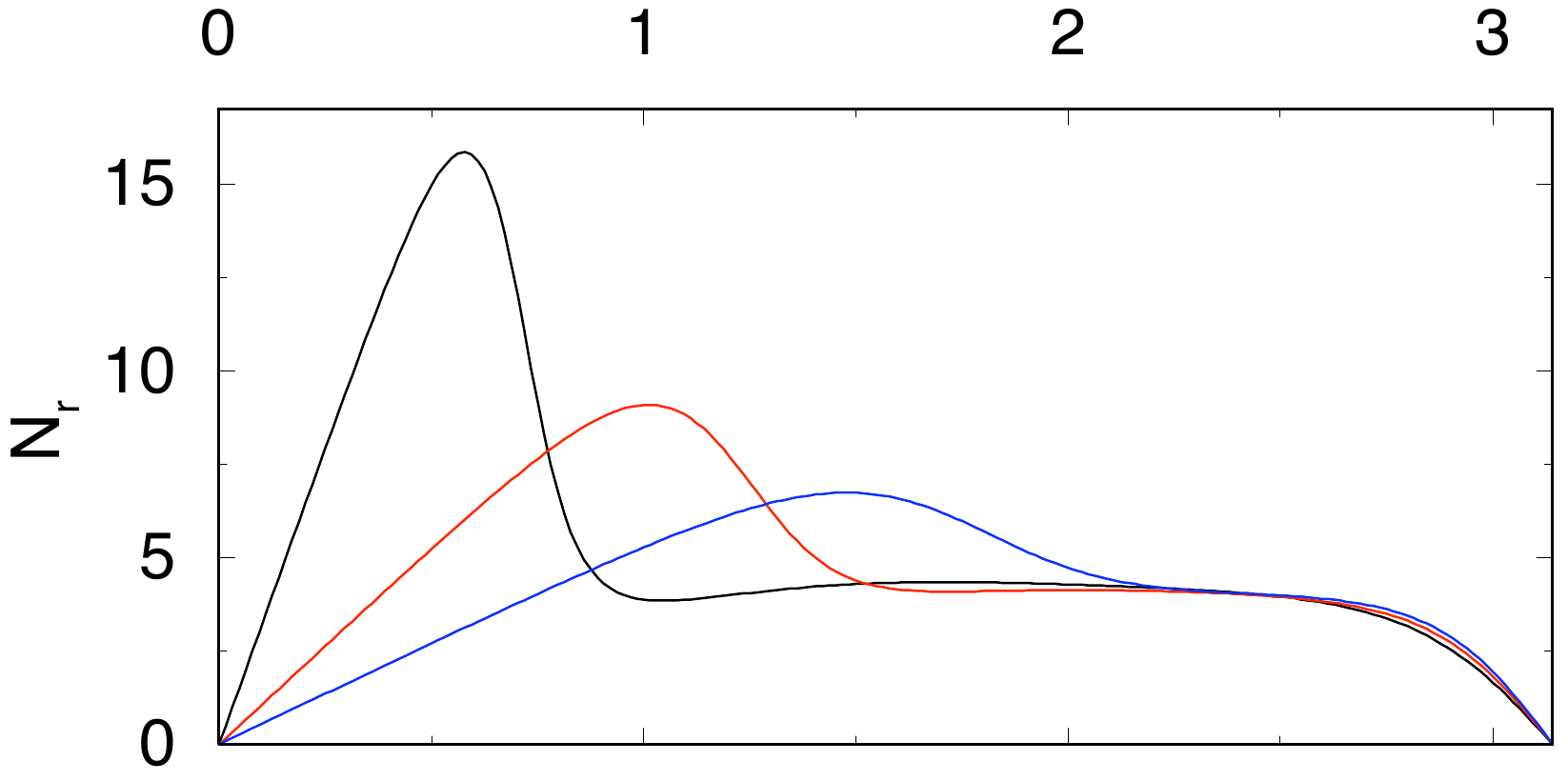}\\
\phantom{....} \ \includegraphics[width=.387\textwidth]{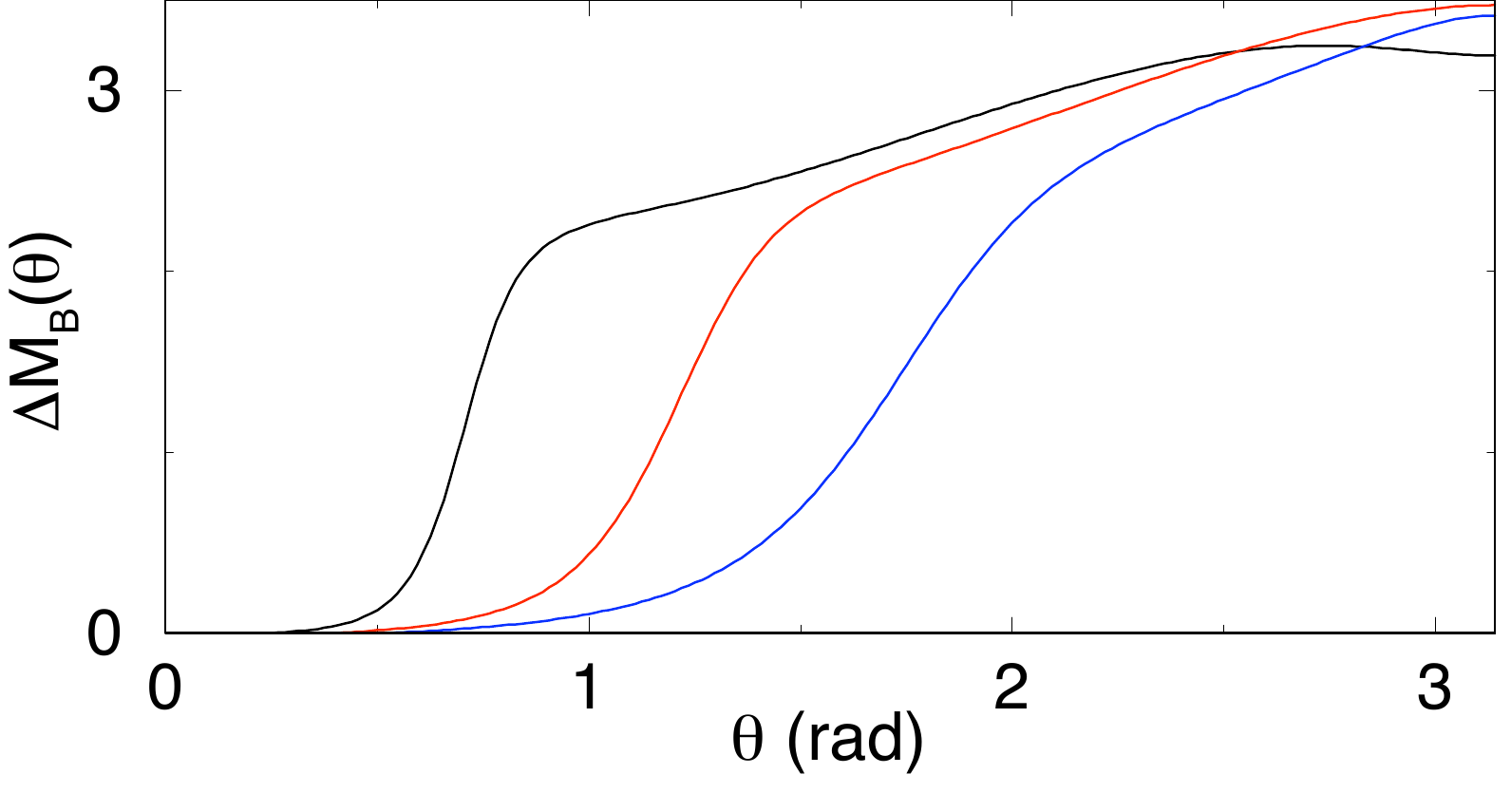}
\end{center}
\caption{Torque $N_r $ (top) and excess chain melting $\Delta M_B$ (bottom) as a function of $\theta$, in 2D. Same parameter values as in Fig.~2B of the body of the text. From left to right, $r=5,15,25$~bp. The elasticity is linear until a threshold $\theta_c\propto \sqrt{r}$, where excessive bending induces bp melting.
\label{bending:fig2D}}
\end{figure}

Following Wiggins \emph{et al.}~\cite{WigginsPRE}, we now derive the average bending moment $N_r$ as a function of the deflection angle $\theta$. In the 2D case, 
\begin{equation}
N_r \equiv \frac{\partial}{\partial \theta} \ln Z(\theta)
\end{equation}
(in units of $k_B T$) measures the torque perpendicular to the substrate that must be applied to two interior monomers separated by $r$ bp in order to impose a deflection angle $\theta$ between them. Examples of $N_r$ vs $\theta$ plots for our model are given in Fig.SI.~\ref{bending:fig2D}. Four regimes appear in these plots:  
\begin{enumerate}
\item[(i)] a linear one at low deflection angle $\theta<\theta_c$; 
\item[(ii)] a non-linear one for intermediate angles; 
\item[(iii)] a saturation plateau for large deflection angles and 
\item[(iv)] a decreasing one near $\theta=\pi$. 
\end{enumerate}
In regime (i), the DNA response is linear, with a GSW  bending moment $N_r=\beta \partial F(\theta,r)/\partial \theta=(\xi^p/r) \theta$, determined only by $\xi^p\simeq \beta\kappa_U$, since melted bps are essentially inexistent.
In the intermediate region (ii), a non-linear behavior occurs. Indeed, it becomes more favorable for the system to break bps in order to make them more flexible and thereby relax the high bending constraint. The plateau appearing in region (iii) shows non-zero response due to the finite value of $\kappa_B$, contrary to the kink model~\cite{WigginsPRE}.
As already mentioned in this Ref.~\cite{WigginsPRE}, when $\theta$ approaches $\pi$ [regime (iv)], $N_r$ vanishes: the symmetry of the system through the axis defined by the vectors $\mathbf{t}_i$ and $\mathbf{t}_{i+r}$ imposes that $\theta=\pi$ is an (unstable) equilibrium point. Indeed, $\theta\in(-\pi,\pi]$ in our calculations, thus larger angles are brought back in this interval modulo $2\pi$. For a given $\theta$, the summation in $Z$ is in fact a sum over all the $\theta+2k\pi$, $k \in \mathbf{Z}$, as illustrated in Fig.SI.~\ref{exs}. Since the contributions of $\theta=\pi$ and $\theta=-\pi$ cancel, $N_r$ vanishes at $\theta=\pi$.  More importantly, we cannot give quantitative answers for cyclization experiments when $|\theta| \approx 2\pi$.

To render these physical mechanisms more explicit, we have also displayed in Fig.SI.~\ref{bending:fig2D} the excess  chain melting $\Delta M_B$ as a  function of $\theta$ (Fig.~2D in the body of the text). It measures the average excess of melted bps in the bended chain as compared to the free, unconstrained one and is given by $\Delta M_B \equiv -\frac{k_BT}2 \frac{\partial}{\partial \mu} \ln p(\theta)$.
We thus confirm that the collapse of $N_r$ corresponds to the proliferation of melted bps.
The typical angle $\theta_c$ at which bending-induced melting occurs is again estimated by equating the energetic cost of bending the polymer in its unmelted state, of order $\beta\kappa_U \theta_c^2/2r$, with the free-energy cost of nucleating a single denaturation bubble (of one bp), $\Delta G_B=4J_0+2K_0+2\mu$~\cite{PMD2}. Again, this argument that leads to $\theta_c\sim \sqrt{r}$  gives a good estimate of the observed threshold. It also gives the upper limit of $r$ for which these non-linearities are apparent, $r_\mathrm{max}\approx \frac{\pi^2}2\kappa_U/\Delta G_B\simeq120$~bp in 2D.

An interesting feature of these calculations is the saturation of $\Delta M_B$ at a finite value [regions (iii) and (iv)], even when $r<r_{\max}$ increases. In Fig.SI.~\ref{bending:fig2D}, this value is close to 3 and the total excess number of denatured bps does not exceed 3 on average. This is corroborated by the fact that the large $\theta$ torque is independent of $r$, mainly due to the few melted bps. In other words, even if $r$~bp, or more, can in principle be melted to relax the bending stress, only a few of them actually do, since it costs more energy to melt more bases, whereas, owing to the small value of $\kappa_B$, a small denaturation bubble suffices to give the whole molecule a very small resistance to torque.

Note that $N_r$ is essentially unchanged between the different fitted parameter sets discussed above, and $\Delta M_B$ varies by at most 20\%, in the melted region only.

\vspace{1cm}

\noindent\textbf{G. Bending-induced melting in 3D}\\

\begin{figure}[ht]
\begin{center}
\includegraphics[width=.4\textwidth]{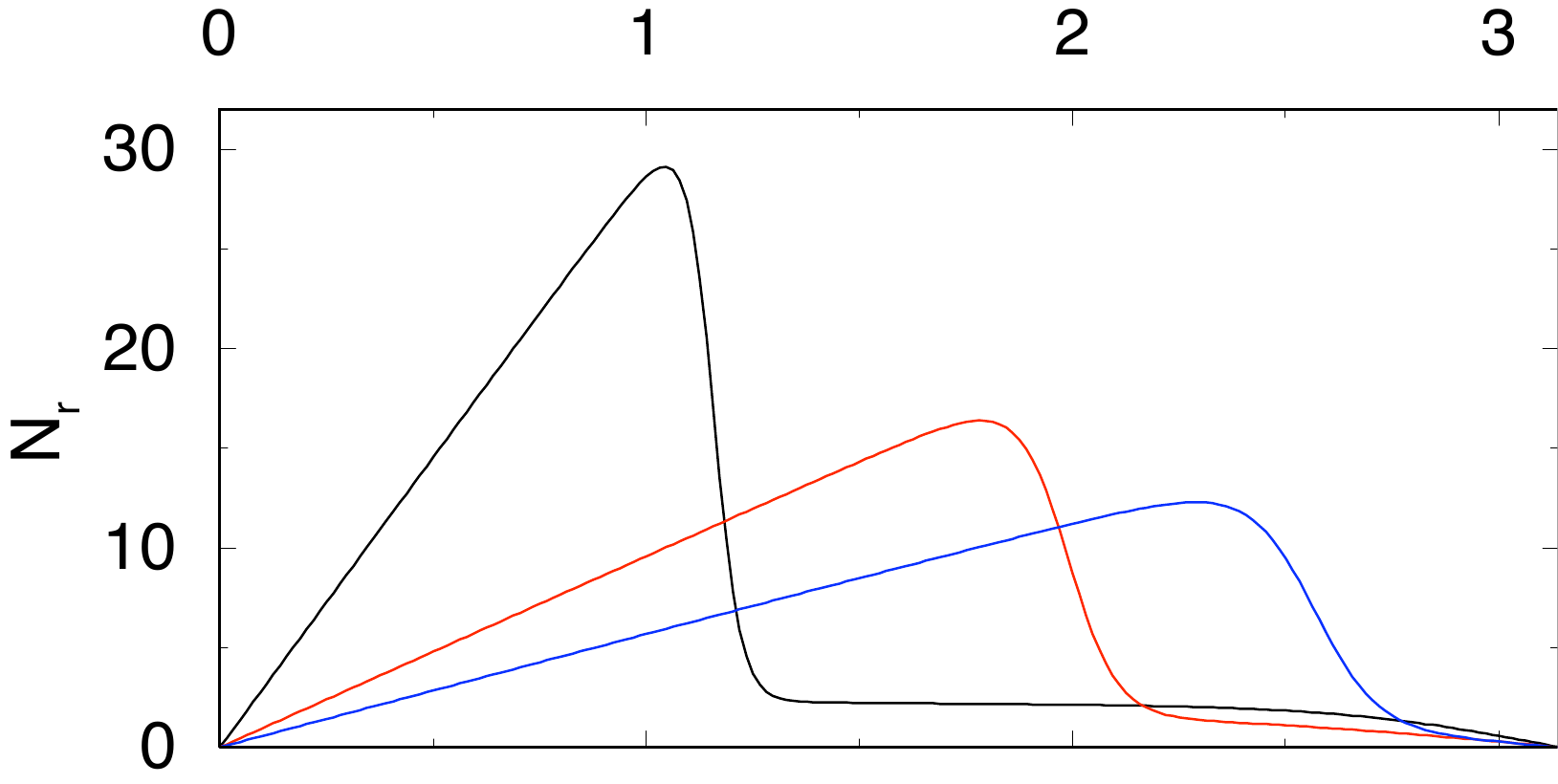}\\
\phantom{..} \includegraphics[width=.4\textwidth]{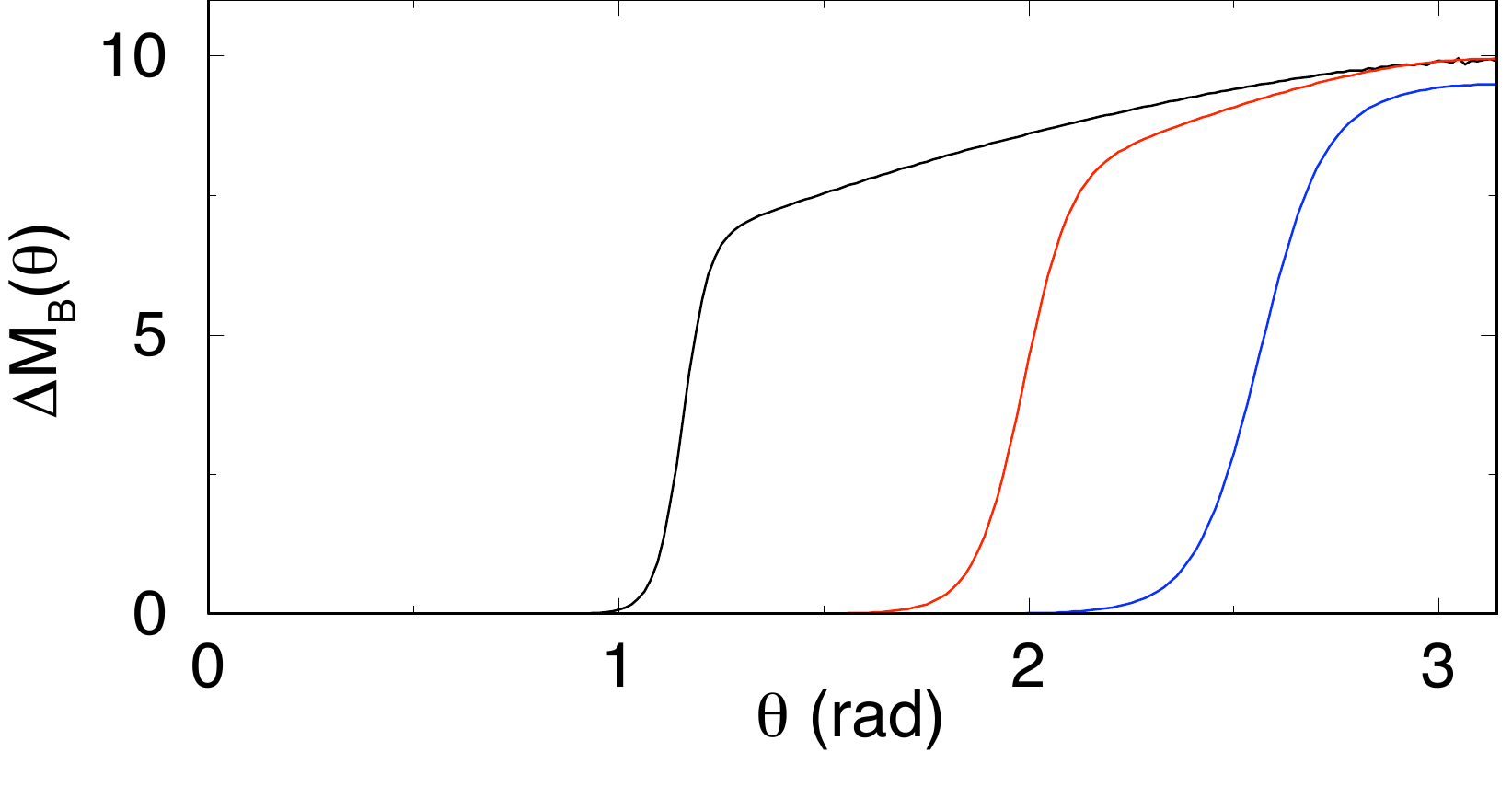}
\end{center}
\caption{Torque $N_r$ (top) and excess chain melting $\Delta M_B$
(bottom) as a function of $\theta$, in 3D. Same parameter values as
in Fig.~2A of the body of the text. From left to right, $r=5,15,25$. The elasticity
is linear until a threshold $\theta_c\propto \sqrt{r}$, where
excessive bending induces bp melting. \label{bending:fig}}
\end{figure}

The previous calculations can be extended in a straightforward way to the 3D case, with very similar qualitative conclusions. The physical meaning of the bending moment is less direct because of the axial symmetry ($m=0$) imposed in the calculation of $p(\cos \theta)$. By contrast, the excess melting, $\Delta M_B$, is meaningful for circular DNAs, where $\theta$ is close to $2\pi$, with the reserves given above. The above argument now leads to $r_\mathrm{max}\approx 50$~bp. In Fig.SI.~\ref{bending:fig} one actually sees that for $r<50$~bp, $\Delta M_B$ saturates at 10~bp near $\theta=\pi$.  In the topical case where $r$ is comparable to the chain persistence length ($\sim 150$~bp), bending-induced melting of constrained DNA is expected to be virtually inexistent in 3D. As for large looped complexes, such as in Ref.~\cite{Pouget} where $r \sim 1000$~bp, melting is not expected to stabilize or to facilitate looping either.

In contrast, non-linear behavior can play a significant role by making DNA much more flexible when the polymer is highly bent, such as in short circular DNA (see~\cite{Lankas}) or in protein-DNA complexes. Our calculations on bending-induced melting, $\Delta M_B$, give a good indication of the degree of melting following a sharp bending constraint. Note, however, that a complete treatment of cyclization would require imposing not only the angle $\theta$ but also the physical distance between bps $i$ and $i+r$~\cite{Yan}. 
Although the full calculation is outside the scope of the present work, imposing $\theta$ does already contain some of the important physical features of cyclization. It is reasonable to expect that even for $\theta\simeq\pi$, our calculation reproduces the correct order of magnitude for cycled DNA. For instance, our 3D predictions concerning excess melting, e.g. $\Delta M_B\approx10$~bp, could be checked by doing UV absorbance measurements on short circular DNAs.

\end{document}